\documentclass{icrc2009}

\usepackage{graphicx}   % for including figures
\usepackage[caption=false]{caption}    % for captions
\usepackage[font=footnotesize]{subfig} % subfig.sty for a double column floating figure using two subfigures
\usepackage{fixltx2e}
\usepackage{url}

\newcommand{\shorttitle}[1]%
{\markboth{Proceedings of the 31\MakeLowercase{$^{st}$} ICRC, {\L}\'{o}d\'{z} 2009}{#1} }
\newcommand{\etal}{\MakeLowercase{\textit{et al. }}} % "et al."

\usepackage{amsmath}
\usepackage{amssymb}

\newcommand{\nhi}{N(\mathrm{H\,\scriptstyle{I}})}

\newcommand{\nhd}{N(\mathrm{H_2})}
\newcommand{\wco}{W_\mathrm{CO}}
\newcommand{\hi}{\mathrm{H\,\scriptstyle{I}}}
\newcommand{\hii}{\mathrm{H\,\scriptstyle{II}}}
\newcommand{\hd}{\mathrm{H}_2}
\newcommand{\xco}{X_\mathrm{CO}}
\newcommand{\ebv}{\mathrm{E[(B-V)]}}
\newcommand{\ebvres}{\ebv_\mathrm{res}}

\begin{document}
\title{\emph{Fermi} observations of Cassiopeia and Cepheus: gamma-ray diffuse emission in the outer Galaxy}

\author{\IEEEauthorblockN{Luigi~Tibaldo\IEEEauthorrefmark{1}\IEEEauthorrefmark{2}\IEEEauthorrefmark{3},
			  Isabelle~A.~Grenier\IEEEauthorrefmark{3}
                          on behalf of the \emph{Fermi} LAT Collaboration}
                            \\
\IEEEauthorblockA{\IEEEauthorrefmark{1} Istituto Nazionale di Fisica Nucleare, Sezione di Padova, 35131 Padova, Italy}
\IEEEauthorblockA{\IEEEauthorrefmark{2} Dipartimento di Fisica ``G. Galilei'', Universit\`a di Padova, 35131 Padova, Italy}
\IEEEauthorblockA{\IEEEauthorrefmark{3} Laboratoire AIM, CEA-IRFU/CNRS/Universit\'e Paris Diderot, Service d'Astrophysique, CEA Saclay,\\ 91191 Gif sur Yvette, France}
}

% please write the preseter's name and short title (3-4 words maximum)
%    which will appear at the header of the even pages.
\shorttitle{L. Tibaldo \etal{} Cas and Cep with \emph{Fermi}}
\maketitle

%\linenumbers

\begin{abstract}
We have used measurements by the \emph{Fermi} Large Area Telescope (LAT) to study the interstellar gamma-ray emission in a region of the second Galactic quadrant, at $\boldsymbol{100^\circ<l<145^\circ}$ and $\boldsymbol{ -15^\circ<b<30^\circ}$. This region encompasses the prominent Gould-Belt clouds of Cassiopeia, Cepheus, and the Polaris flare, as well as large atomic and molecular complexes at larger distances in several spiral arms. The good kinematic separation in velocity between the local, Perseus, and outer arms, and the presence of massive complexes in each, make this region very well suited to probe the gamma-ray emission from the interstellar medium beyond the solar circle. The unprecedented sensitivity and angular resolution of the LAT provide improved constraints on the gradient of the cosmic-ray densities and on the increase of the CO-to-$\boldsymbol{\hd}$ conversion factor, $\boldsymbol{\xco}$, in the outer Galaxy.
\end{abstract}


\section{Introduction}
Galactic interstellar gamma-ray emission is produced through the interactions of cosmic rays (CR) with the interstellar gas (via $\pi^0$ production and Bremsstrahlung) and with the interstellar radiation field (via Inverse Compton interactions). Thus gamma rays can be used to trace cosmic rays in the Galaxy and the properties of the interstellar medium (ISM), in particular the total gas column densities.

The interpretation of the observed gamma-ray diffuse emission is often based on two radio tracers of the interstellar gas. The 21 cm line of the hyperfine transition of atomic hydrogen $\hi$ is used to derive its column density $\nhi$. The integrated 2.6 mm line intensity of the rotational transition of CO, $\wco$, is assumed to be proportional to the column density of molecular hydrogen $\nhd$ through the conversion factor $\xco$, $\xco = \nhd/\wco$. The estimate of the $\xco$ ratio is necessary for measuring $\hd$ cloud masses from radio observations and it also impacts the derivation of the distribution of Galactic cosmic-ray sources from gamma-ray observations \cite{strongXvar}, but our current knowledge is limited, especially beyond the solar circle. Similarly, the gradient of the gamma-ray emissivity of the more diffuse $\hi$ gas provides useful constraints on the gradient of the cosmic-ray densities in the outer Galaxy, but its derivation was so far limited by the sensitivity and angular resolution of gamma-ray telescopes.

The situation has greatly improved on both grounds with the successful launch of the \emph{Fermi} Gamma-ray Space Telescope on June 11, 2008. The Large Area Telescope (LAT) on board the \emph{Fermi} mission \cite{latpaper} detects photons from 20 MeV to $>\! 300$ GeV. The sensitivity of the LAT is more than an order of magnitude greater with respect to the previous EGRET telescope on board the Compton Gamma-Ray Observatory; in addition, it has a superior single photon angular resolution (e.g., the 68\% containement angle for 1 GeV photons at normal incidence reaches $0.6^\circ$ for the LAT with respect to $1.3^\circ$ for EGRET).

Here we present the analysis of the diffuse gamma-ray emission observed by the \emph{Fermi} LAT in a selected region of the second Galactic quadrant, at $100^\circ <l < 145^\circ$ and $-15^\circ < b < 30^\circ$, during the first 9 months of the science phase of the mission.

\section{Modelling of gamma-ray emission}
Our aim is to separate the gamma-ray emission from four regions along the line-of-sight in the second Galactic quadrant:
\begin{enumerate}
\item the very nearby gas complexes in the Gould Belt, at hundreds of parsecs from the solar system \cite{gouldbelt};
\item the main part of the local arm, typically 1 kpc away;
\item the Perseus arm, about 4 kpc away;
\item the outer arm and beyond.
\end{enumerate}
In the $(l,b)$ interval we have chosen the division is possible due to the good kinematic separation of the $\hi$ and CO lines along the lines of sight and the low spatial degeneracy between the resulting maps in each region and gas phase.

To estimate the gas column densities in each region we derived maps from the LAB survey \cite{LABsurvey} for $\hi$ and from the moment-masked composite survey by Dame \etal \cite{COsurvey} for CO. The following separation scheme was developed and applied to separate the four regions in velocity. The preparation of the gas maps started from preliminary velocity boundaries given in terms of Galactocentric annuli ($R < 8.8$ kpc, $8.8$ kpc $< R < 10$ kpc, $10$ kpc $< R < 14$ kpc, and $R > 14$ kpc) using a flat rotation curve with $R_\odot=8.5$ kpc and a rotation velocity of 220 km s$^{-1}$. The velocity boundaries were adjusted for each line of sight to better separate the different cloud structures based on their coherence in the $(l,b,v)$ phase space. These boundaries were used to produce $\nhi$ and $\wco$ maps in each region. The broad $\hi$ lines can spill over from one velocity band into the next. To correct for this cross-contamination between adjacent regions, each $\hi$ spectrum was fitted by a combination of Gaussian lines. The latter were used to correct the integral in a particular velocity band from the spill-over from adjacent lines. The $\hi$ column densities were derived applying the optical depth correction under the usual approximation of a uniform spin temperature $T_s = 125$ K.

An additional map was prepared to account for the so-called dark gas, i. e. neutral interstellar gas not traced by $\hi$ or CO. Grenier \etal \cite{darkclouds} reported a considerable amount of this gas at the interface between the phases traced by $\hi$ and CO. The dark-gas map was derived from the $\ebv$ map by Schlegel \etal \cite{ebv}, which traces the total dust column density in our Galaxy. This reddening map was fitted with a linear combination of the $\nhi$ and $\wco$ maps previously described, to subtract the parts linearly correlated with this set of maps. The residual $\ebvres$ map traces the distribution of the gas column density not traced by $\hi$ and CO.

The Inverse Compton (IC) emission expected over this narrow part of the sky is rather uniform and so indistinguishable from the isotropic background (the mixture of truly isotropic gamma-ray emission, unresolved point sources and residual background from misclassified interactions of cosmic rays in the LAT). Therefore we inserted only an isotropic term in the analysis model, to account for all of them.

In this preliminary analysis we neglect any contribution from the ionized hydrogen $\hii$. Thus, assuming that the ISM is transparent to gamma rays, that the characteristic diffusion lengths for cosmic electrons and protons exceed the dimensions of cloud complexes, and that cosmic rays penetrate clouds uniformly to their core, we can model the gamma-ray flux $\Phi$ in a direction $(l,b)$ by equation \ref{anamodel}:
\begin{eqnarray}
\Phi(l,b) & = & \sum_\imath \left[ A_\imath \cdot \nhi(l,b)_\imath + B_\imath \cdot \wco(l,b)_\imath \right] + \nonumber \\
&+& C +\sum_\jmath D_\jmath \cdot \delta^{(2)}(l-l_\jmath,b-b_\jmath) + \nonumber \\
&+& E \cdot \ebvres(l,b)\label{anamodel}
\end{eqnarray}
The sum over $\imath$ represents the combination of the four different Galactic regions. The parameters are the emissivity of $\hi$, $A_\imath$, and the emissivity per unit of CO emission intensity, $B_\imath$. The isotropic term, $C$, accounts for all the components discussed above, including IC emission. The contribution from the 9 point sources found in the region in the LAT Bright Source List \cite{BSlist} at a significance level above $10 \, \sigma$  is represented by the sum over $\jmath$. For point sources we used the published positions, power-law spectra were fitted independently in each energy range.  Three other point sources within $5^\circ$ outside of the region boundaries were included, keeping their spectra as determined in the 6 month catalog internally available to the LAT Collaboration. The $E$ parameter is the emissivity per unit of $\ebvres$.

\section{Gamma-ray analysis}

The gamma-ray data were obtained by the \emph{Fermi} LAT during the period August 4 2008 - April 26 2009. The standard event selection was applied, selecting the \emph{Diffuse} event class (the class with the least residual instrumental background),  and rejecting events seen at a zenith angle greater than $100^\circ$ and time intervals where the rocking angle respecting to zenith was greater than $39^\circ$, in order to limit the contamination from Earth albedo gamma rays.

%figure for the next section
\begin{figure*}[!t]
\centerline{
\subfloat[0.3 -- 0.6 GeV]{\includegraphics[width=0.5\textwidth]{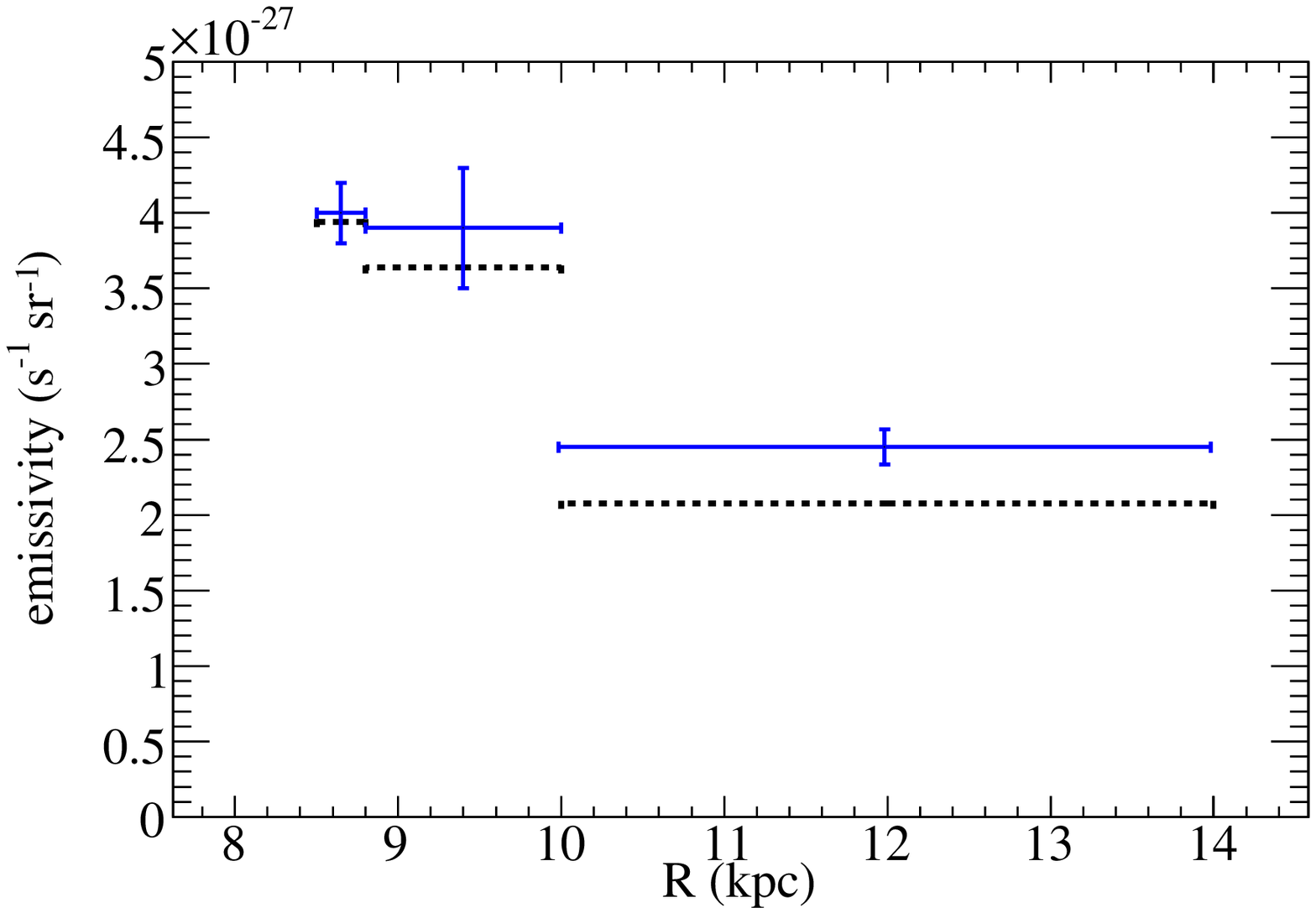}\label{a}}
\hfil
\subfloat[0.6 -- 1 GeV]{\includegraphics[width=0.5\textwidth]{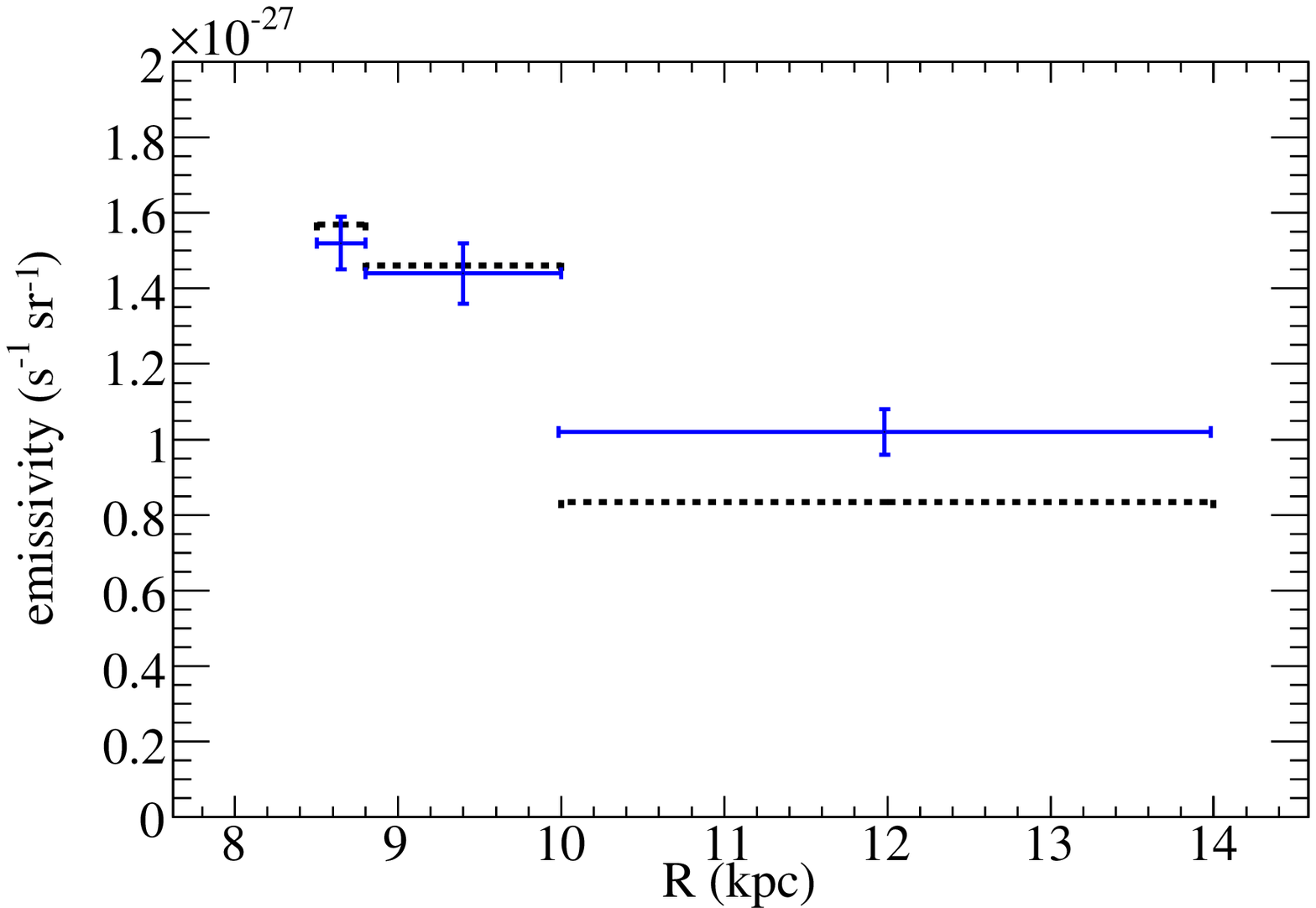}\label{b}}
}
\centerline{
\subfloat[1 -- 2 GeV]{\includegraphics[width=0.5\textwidth]{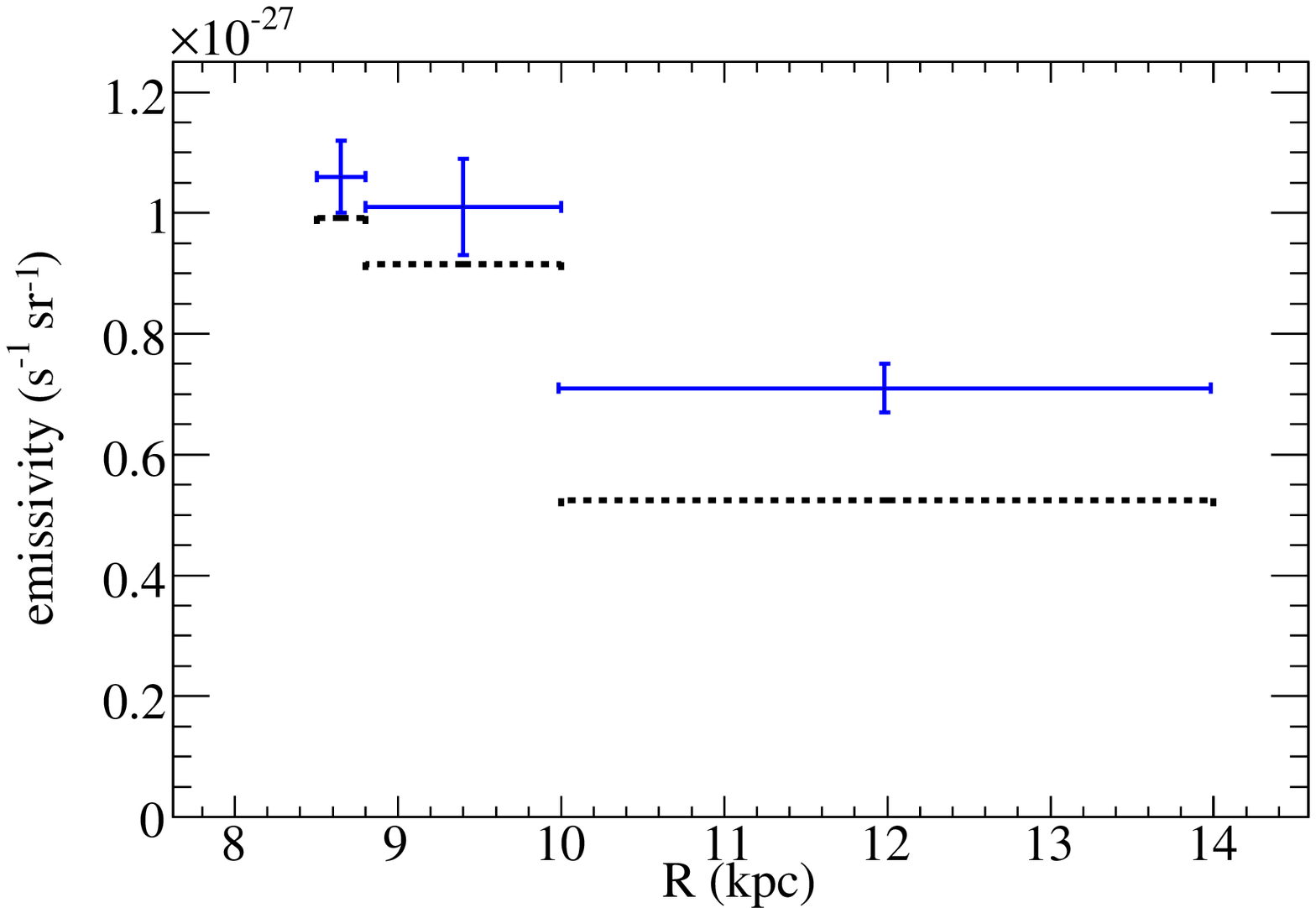}\label{c}}
\hfil
\subfloat[2 -- 10 GeV]{\includegraphics[width=0.5\textwidth]{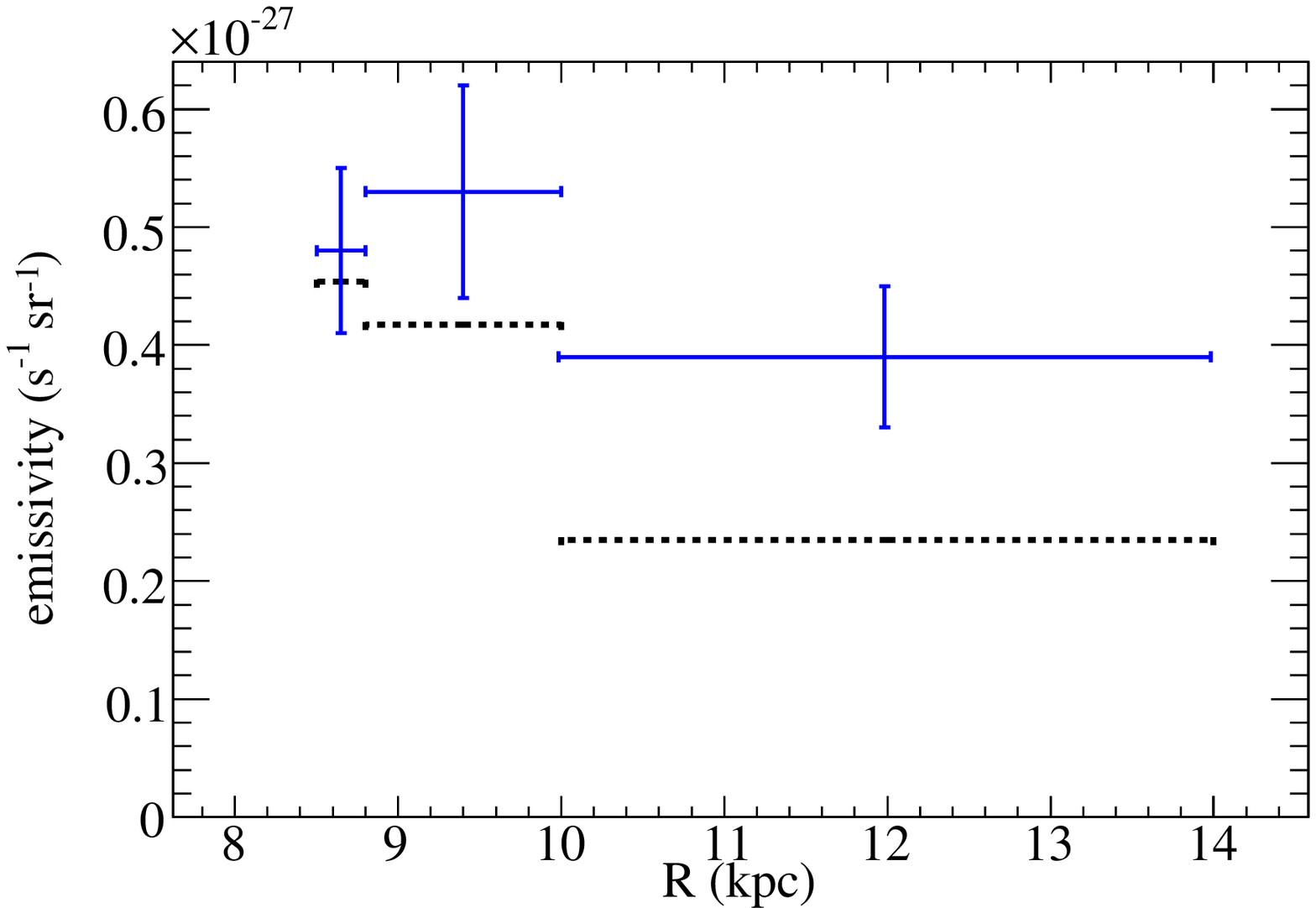}\label{d}}
}
\caption{(PRELIMINARY) Dependence on Galactocentric radius $R$ of the $\hi$ emissivity measured in the four energy bands: a) 0.3~--~0.6 GeV, b) 0.6~--~1 GeV, c) 1~--~2 GeV, d) 2~--~10 GeV. Black dashed lines give the predictions from the GALPROP model \texttt{54\_77Xvarh7S}, i.e. the sum of the gas emissivity due to $\pi^0$ decay and Bremsstrahlung, scaled by $+15$\%. The three regions (blue crosses) spanning different ranges in $R$ encompass the Gould Belt, the local arm and the Perseus arm (from left to right); the error bars include systematics due to the uncertainties on the event selection efficiency.}\label{emissprofile}
\end{figure*}
%end of following figure

The LAT data were analysed using a binned maximum Likelihood procedure with Poisson statistics, on a spatial grid with $0.5^\circ$ spacing using the LAT \emph{Science Tools}\footnote{http://fermi.gsfc.nasa.gov/ssc/data/analysis/documentation/Cicerone/}. The analysis was done for four contiguous energy bands: 0.3~--~0.6 GeV, 0.6~--~1 GeV, 1~--~2 GeV and 2~--~10 GeV. A power-law spectrum was fitted independently for each diffuse component  in each energy range, allowing a careful calculation of the exposures. Over such a narrow region of the sky we were able to perform a full convolution with the energy-dependent PSF. The analysis used the \texttt{P6\_V3} Instrument Response Functions; with respect to pre-flight Response \cite{latpaper} they include inflight calibration and account for the loss of efficiency due to pile-up events in the LAT \cite{perfpost}.

We verified that the use of an IC emission map derived from the GALPROP cosmic-ray propagation code \cite{galprop98,galprop04} does not impact the resulting gas emissivities. The effect of the point sources was assessed by adding one by one the sources in order of decreasing detection significance: only the three brightest have an impact on the gas emissivities, except for the outer arm. Due to the small amount of gas in this region and the lack of significant features in its spatial distribution, the gas emissivity responds to the presence of point sources along the Galactic plane. Thus the current $A_4$ value must be regarded only as an upper limit and will not be used in the following discussion.

A detailed description of the analysis, including a careful treatment of the systematics, will be given in a paper in preparation within the \emph{Fermi} LAT Collaboration. In the following section we discuss some preliminary results.

\section{Results}

\subsection{$\hi$ emissivities}

The integrated $\hi$ emissivities measured in the four energy bands are given in Fig.~\ref{emissprofile}. The present values in the Gould Belt are consistent within 10\% with the local $\hi$ emissivity spectrum derived from LAT data in a mid-latitude region of the third Galactic quadrant \cite{hiemiss}.

The physical interpretation of the emissivities is based on the comparison with a GALPROP model. GALPROP is a numerical code for cosmic-ray propagation in the Galaxy, see e.g. \cite{galprop98,galprop04}, widely used to analyse LAT data. The program has recently been updated with the inclusion of new astronomical surveys, more accurate description of the interstellar gas, improvements in the modelling of physical processes, the cosmic-ray electron spectrum derived from \emph{Fermi} data \cite{Fermie} and tuning to the diffuse gamma-ray emission seen by \emph{Fermi} itself \cite{andytalk}. The GALPROP run used for this work, \texttt{54\_77Xvarh7S}, is consistent with directly measured CR spectra and the non-confirmation of the EGRET GeV excess \cite{GeVexc, icrcGeVexc}. It has been scaled by $+15$\% in Fig.~\ref{emissprofile} to provide a straightforward comparison of the emissivity decrease towards the outer Galaxy. This factor is the average of the data-over-GALPROP ratios found in the four energy bands for the Gould Belt.

Provided this small intensity normalization, the consistency between the data and GALPROP points in the Gould Belt at all energies shows that the gamma-ray emissivity of local $\hi$ can be explained by the cosmic-ray spectrum measured at Earth. This confirms for very nearby gas the conclusion reached from the analysis of LAT data over broader regions of the sky \cite{GeVexc, icrcGeVexc}: LAT observations are inconsistent with the GeV excess seen by EGRET over the whole sky and in specific local $\hi$ complexes \cite{egretmono}. No significant variations in emissivity are found between the Gould Belt and the local arm, as can be expected from the large coupling length of $\sim 1.75$ kpc found between cosmic protons and interstellar gas \cite{egretdiff}.

The emissivity profiles in Fig.~\ref{emissprofile} significantly decrease with Galactocentric distance, as expected from the distribution of candidate cosmic-ray sources used in the GALPROP model. However, the measured decrease is found to be systematically smaller from the local to the Perseus arm than the GALPROP prediction, especially at high energies where the higher angular resolution allows a better component separation (Fig.~\ref{c} and \ref{d}).

\subsection{Calibration of molecular masses}

\begin{figure}[!t]
\begin{center}
\includegraphics[width=0.5\textwidth]{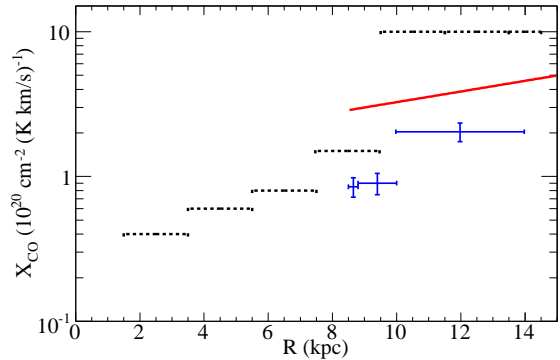}
\caption{(PRELIMINARY) Variation in function of Galactocentric radius of the $\xco$ conversion factor. Blue crosses are our measurements in the distance ranges corresponding to the Gould Belt, the local arm and the Perseus arm (from left to right). Dashed black lines represent the values used by Strong \etal \cite{strongXvar} in GALPROP. The solid red line is the conversion function determined for the outer Galaxy by Nakanishi \& Sofue \cite{Xjap} from CO data and virial masses (adapted to the rotation curve assumed for our analysis).}\label{xcofig}
\end{center}
\end{figure}

The emissivities can be used to estimate the CO-to-$\hd$ conversion factor. In each energy range and each region we can derive $\xco{}_\imath=B_\imath /2 A_\imath$. Derived as a flux ratio between two components with the same spectral shape, the $\xco$ factor is not affected by systematic uncertainties in the instrument efficiency. It is also insensitive to the presence of isolated point sources not included in the analysis, because the large angular extent and characteristic shapes of the clouds provide better discrimination in the fitting procedure, thanks to the good angular resolution of the LAT. However, the $\xco$ factor is sensitive to the inclusion of the dark gas in the model. The impact is not highly significant in the very nearby Gould-Belt clouds, where we have ample resolution and little degeneracy between the $\hi$, dark, and CO phases. But, since the three phases blend more when seen from a larger distance, the separation is less effective and the $\xco$ measurement in the more distant complexes becomes more sensitive to the inclusion of the dark gas. For instance, the $\xco$ estimate in the local and Perseus arms can increase by 30\% if the dark gas is not included in the analysis. The estimate of $\xco$ is also sensitive to the level of degeneracy between $\nhi$ and $\wco$ maps that is inherent to the physical structure of the clouds. Since the angular resolution of the LAT is strongly energy-dependent, this can lead to an energy-dependence of the $\xco$ measurement for distant clouds. The small spread of $\xco$ measurements with energy can be considered as a systematic uncertainty on the component separation. The results are summarized in Fig.~\ref{xcofig}, where $\xco$  is evaluated for each region as the weighted average of the values obtained in the four energy ranges and the error bars include these systematic uncertainties.

The value measured in the Gould-Belt clouds agrees with the results obtained by EGRET for the Cepheus and Polaris flares \cite{egretcep}, and is consistent with estimates for other clouds in the solar neighborhood \cite{egretmono}. The Gould-Belt clouds are within 300 pc from the solar system, so, given the $\lesssim 0.6^\circ$ angular resolution of the LAT above 1 GeV, they are mapped in gamma-rays with a spatial accuracy $\lesssim 3$ pc. They also show a very low attitude at forming massive stars. Therefore it is very unlikely that the corresponding $\xco$ measurement is biased by a population of unresolved sources closely following the structure of the clouds. This bias is unlikely too in the local-arm region, but source contamination cannot be ruled out for the $\xco$ measurement in the more remote Perseus arm.

The present results suggest an increase of $\xco$ in the outer Galaxy, as already deduced by many other observations (see e.g. Sodroski \etal \cite{sodroski} and the other references cited in \cite{strongXvar}). Whether the present $\xco$ gradient can be fully attributed to the metallicity gradient, or partially to unresolved sources, $\hi$ and CO separation problems, or gas not traced by $\hi$ and CO, needs further investigation. In any case, the results shown in Fig.~\ref{xcofig} indicate smaller values of $\xco$ with respect to those used by Strong \etal \cite{strongXvar} in GALPROP: the present measurement is significantly lower in the outer Galaxy. The values preliminarily obtained from LAT data are also systematically smaller than the $\xco(R)$ relation determined from CO data and virial masses beyond the solar circle \cite{arimoto96,Xjap}.

\section*{Acknowledgments}

The \emph{Fermi} LAT Collaboration acknowledges support from a number of agencies and institutes for both development and the operation of the LAT as well as scientific data analysis. These include NASA and DOE in the United States, CEA/Irfu and IN2P3/CNRS in France, ASI and INFN in Italy, MEXT, KEK, and JAXA in Japan, and the K.~A.~Wallenberg Foundation, the Swedish Research Council and the National Space Board in Sweden. Additional support from INAF in Italy for science analysis during the operations phase is also gratefully acknowledged.

\end{document}